\documentclass[final]{raa06}           
\usepackage{graphicx,times}             
\usepackage{natbib}
\usepackage{amssymb,amsmath}
\bibpunct{(}{)}{;}{a}{}{,}
\usepackage[colorlinks=true, citecolor=blue]{hyperref}%

\usepackage{graphicx,kantlipsum,setspace}
\usepackage{caption}
\usepackage{graphics,epsf}
\usepackage{amsmath}                
\usepackage{amsfonts}               
\usepackage{amssymb}                
\usepackage{epsfig}                 
\usepackage{appendix}
\usepackage{graphicx}
\usepackage{float}
\usepackage{color}
\usepackage{multirow}
\usepackage{colortbl}
\usepackage[para,online,flushleft]{threeparttable}
\usepackage{xcolor}

\hypersetup{citecolor=blue, 
            linkcolor=red, 
            menucolor=blue, 
            urlcolor=blue}  

\newcommand{\yr}{{~\rm yr}}

\newcommand{\pc}{{~\rm pc}}

\newcommand{\keV}{{~\rm keV}}

\begin{document}

   \title{Point-symmetry in SNR G1.9+0.3: A supernova that destroyed its planetary nebula progenitor
}

   \volnopage{Vol.0 (20xx) No.0, 000--000}      
   \setcounter{page}{1}          

   \author{Noam Soker
    }

   \institute{Department of Physics, Technion, Haifa, 3200003, Israel;   {\it   soker@physics.technion.ac.il}\\
\vs\no
   {\small Received~~20xx month day; accepted~~20xx~~month day}}

\abstract{I analyze a new X-ray image of the youngest supernova remnant (SNR) in the Galaxy, which is the type Ia SNR G1.9+0.3, and reveal a very clear point-symmetrical structure. Since explosion models of type Ia supernovae (SNe Ia) do not form such morphologies, the point-symmetrical morphology must come from the circumstellar material (CSM) into which the ejecta expands. The large-scale point-symmetry that I identify and the known substantial deceleration of the ejecta of SNR G1.9+0.3 suggest a relatively massive CSM of $\gtrsim 1 M_\odot$. I argue that the most likely explanation is the explosion of this SN Ia into a planetary nebula (PN). The scenario that predicts a large fraction of SN Ia inside PNe (SNIPs) is the core degenerate scenario. Other SN Ia scenarios might lead to only a very small fraction of SNIPs or not at all.  
 }

 \authorrunning{N. Soker}            
\titlerunning{The point-symmetric structure of SN 1987A}  
   
      \maketitle

\section{INTRODUCTION}
\label{sec:intro}

A point-symmetric morphology is composed of pairs of twin structural components on opposite sides of the center of the nebula. Such structures are clearly observed in many tens of planetary nebulae (PNe), as catalogues of PNe (and proto-PNe) reveal (e.g.,  \citealt{Balick1987, Chuetal1987, Schwarzetal1992, CorradiSchwarz1995, Manchadoetal1996, SahaiTrauger1998, Sahaietal2011, Parkeretal2016, Parker2022}). Many PNe are shaped by jets (e.g., \citealt{Morris1987, Soker1990AJ, SahaiTrauger1998}), including point-symmetric morphologies (e.g., \citealt{Sahai2000, Sahaietal2007, Sahaietal2011, Boffinetal2012}). Many PNe and proto-PNe (e.g., \citealt{Sahaietal2000, Sahaietal2011}), like the post-asymptotic giant branch (AGB) star IRAS 16594-4656 \citep{Hrivnaketal2001}, show that the point-symmetry is not always perfect. Namely, they might have some deviations from perfect point symmetry. In particular, the two opposite clumps/lobes/arcs/filaments of a pair might have different structures, differ in brightness, be not exactly at $180^\circ$ to each other with respect to the center (bent-morphology), and have different distances from the center. As well, non-paired clumps might exist in the nebula. 
  
{ PNe leave white dwarf (WD) remnants, in many cases a WD in a binary system. If the WD remnant explodes as a type Ia supernova (SN Ia) before the PN is dispersed into the interstellar medium (ISM), the PN might have an imprint on the morphology of the SN remnant (SNR). An SN inside a PN is termed SNIP (e.g., \citealt{TsebrenkoSoker2015SNIP}). } 
Not all theoretical SN Ia and peculiar SN Ia scenarios allow for the formation of point-symmetric SNRs (for some recent reviews of the scenarios, but without reference to point-symmetry, see, e.g., \citealt{Hoeflich2017, LivioMazzali2018, Soker2018Rev, Soker2019Rev, Wang2018,  Jhaetal2019NatAs, RuizLapuente2019, Ruiter2020, Liuetal2023Rev}).
 
The formation process of PNe, { typically tens to thousands of years to form the dense shell, } is much longer than the dynamical time of the AGB progenitor, { about one year. } Also, the launching phase of the jets by a companion to the AGB progenitor is much longer than the dynamic time of the accretion disk that launches the jets. This allows for disk precession that launches opposite pairs of jets in varying directions. In SN Ia scenarios that involve a disk with bipolar explosion morphology (e.g., \citealt{Peretsetal2019, Zenatietal2023}), the disk explosion time is not much longer, and even shorter, than the dynamical time of the disk. No disk precession is possible during the explosion. If a SNR Ia has a point symmetry it seems that it results from a point-symmetric circumstellar material (CSM). 
 
Peculiar SNe Ia might have also peculiar morphologies, such as the unordered morphology of the peculiar SNR 3C 397 (e.g., \citealt{Ohshiroetal2021})  that might result from deflagration (\citealt{Mehtaetal2023}). However, these are not expected to form point-symmetric morphologies. ISM magnetic fields might shape only one pair of twin structural features (e.g., \citealt{Wuetal2019}) and might play other roles in SNRs (e.g., \citealt{Xiaoetal2022}).
\cite{Velazquezetal2023} simulate non-spherical pre-explosion mass loss into a magnetized ISM. They find that when the pre-explosion wind is axisymmetric (rather than spherical) and its symmetry axis is inclined to the ISM magnetic field then the ears in the SNR might be bent. However, point-symmetric clumps/filaments cannot be formed by this mechanism. Surrounding density inhomogeneities might also shape SNRs (e.g., \citealt{Luetal2021}).  However, these ISM effects cannot form point-symmetric structures. \cite{Zhangetal2023} simulated the shaping of SNR G1.9+0.3 with magnetic fields and ISM density gradients. They could form a pair of ears, but not a point-symmetry (which was not known then). \cite{GriffethStoneetal2021} simulated SNR G1.9+0.3 as a highly non-spherical explosion into a uniform medium. This cannot form a point-symmetric structure.  In a recent study \cite{Villagran2023} conduct three-dimensional magneto-hydrodynamic simulations to reproduce the morphology and emission of SNR G1.9+0.3 by a non-spherical pre-explosion wind into a magnetized ISM. They also obtained an axisymmetrical morphology, but not a point-symmetry. { Instabilities that develop in the ejecta-ISM interaction are not expected to form point-symmetric morphologies. Furthermore, \cite{Mandaletal2023} demonstrate with hydrodynamical models that the instabilities that develop as SNRs interact with an ambient medium have a characteristic peak in their power spectra that is relatively large, $>10$. This cannot account for a point-symmetric structure with only a few prominent pairs of opposite morphological features.  } 

In this study, I identify a point-symmetric morphology in the newly released  X-ray image of SNR G1.9+0.3 \citep{Enokiyaetal2023}, a young SNR Ia that exploded around 1890-1900 (e.g., \citealt{Carltonetal2011, Chakrabortietal2016, Borkowskietal2017, Pavlovic2017}). I analyze the image in section \ref{sec:PointSymmetry} and conclude that the most likely explanation is that this SNR was shaped by an SN Ia inside a PN, i.e., an SNIP. 

\cite{TsebrenkoSoker2015G1903} already suggested that SNR G1.9+0.3 is a SNIP, and simulated its shaping. However, they did not refer to point-symmetry. The present analysis put the SNIP suggestion on a very solid ground. To facilitate the analysis and discussion in section \ref{sec:Summary}, I start by considering the ability of different SN Ia scenarios to account for point-symmetric morphologies (section \ref{sec:Scenarios}).

\section{Point symmetry in SN Ia scenarios}
\label{sec:Scenarios}

In Table \ref{Tab:Table1} I list SN Ia scenarios (first row) with some of their properties (second row). The properties are 
the number of stars in the system at the time of explosion $N_{\rm exp}$, the number of surviving stars after the explosion $N_{\rm sur}$, the mass of the exploding white dwarf (WD) where $M_{\rm Ch}$ stands for near Chandrasekhar mass, and the morphology of the ejecta ($E_{\rm j}$) being spherical (S) or non-spherical (N). These properties refer to normal SNe Ia where the WD that explodes does not leave a remnant.  
The first two rows of the table are from a much larger table from \cite{Soker2019Rev} which compares the scenarios with each other and with observations. {  Scenarios where there is only one star at the explosion, $N_{\rm exp}=1$, are grouped into \textit{lonely-WD scenarios}, and might account for most, or even all, normal SNe Ia \citep{BraudoSoker2023}. } 
\begin{table*}
\scriptsize
\begin{center}
  \caption{SN Ia scenarios and their ability to form a point-symmetric SNR}
    \begin{tabular}{| p{2.4cm} | p{1.5cm}| p{1.6cm}| p{1.6cm}| p{1.6cm} | p{1.6cm} | p{1.6cm} |}
\hline  
\textbf{{Scenario$^{[{\rm 1}]}$}}  & {Core Degenerate \newline (CD)}    & {Double Degenerate} (DD) & {Double Degenerate} (DD-MED)& {Double Detonation} (DDet) & {Single Degenerate} (SD-MED) & {WD-WD collision} (WWC)\\
\hline  
 {$\mathbf{[N_{\rm exp}, N_{\rm sur}, M, Ej]}$$^{[{\rm 2}]}$}
  & $[1,0,M_{\rm Ch},{\rm S}]$ 
  & $[2,0,$sub-$M_{\rm Ch},{\rm N}]$
  & $[1,0,M_{\rm Ch}, {\rm S}]$ 
  & $[2,1,$sub-$M_{\rm Ch},{\rm N}]$
  & $[2,1,M_{\rm Ch},{\rm S}]$  
  & $[2,0,$sub-$M_{\rm Ch},{\rm N}]$ \\
\hline  
\textbf{Point symmetry in the SNR} & Expected in some SNIPs with point-symmetric PNe. & Very rare: A SNIP with point-symmetric PN. & Very rare: A SNIP with point-symmetric PN.& Extremely rare. & Possible: a symbiotic progenitor; Low-mass CSM. & { Extremely rare; large-scale departures from an elliptical shape. } \\              
\hline  
%
%
     \end{tabular}
  \label{Tab:Table1}\\
\end{center}
\begin{flushleft}
\small 
Notes: \small [1] Scenarios for SN Ia by alphabetical order. MED: Merger to explosion delay time. It implies that the scenario has a delay time from merger or mass transfer to explosion. MED is an integral part of the CD scenario.   
 \\
 \small [2] $N_{\rm exp}$ is the number of stars in the system at the time of explosion;  $N_{\rm sur}$ is the number of surviving stars in normal SNe Ia: $N_{\rm sur}=0$ if no companion survives the explosion while $N_{\rm sur}=1$ if a companion survives the explosion (in some peculiar SNe Ia the exploding WD is not destroyed and it also leaves a remnant); $M_{\rm Ch}$ indicates a (near) Chandrasekhar-mass explosion while sub-$M_{\rm Ch}$ indicates sub-Chandrasekhar mass explosion; Ej stands for the morphology of the ejecta, where S and N indicate whether the scenario might lead to spherical explosion or cannot, respectively.   
\end{flushleft}
\end{table*}

Here I add to Table \ref{Tab:Table1} the third row that indicates whether the scenario might lead to a point-symmetric SNR, and describe below the scenarios only in their relation to a point-symmetric SNR. 

The \textit{core-degenerate} (CD) scenario predicts that a large fraction of SNe Ia occurs inside PNe or PN remnants. These are term SNIPs for SNe Ia Inside PNe. A PN remnant is an old PN shell that at the time of the explosion is mostly neutral and hence does not shine as a PN. The reason that the CD scenario predicts many SNIPs is that the core and the WD merge during or at the end of the common envelope evolution (CEE; e.g., \citealt{KashiSoker2011, Ilkov2013, AznarSiguanetal2015}), and might explode within several hundreds of thousands of years, which is the merger to explosion delay (MED) time. In \cite{Soker2022Delay} I estimated that the fraction of SNIPs among all normal SNe Ia in the Milky Way and the Magellanic clouds is $f_{\rm SNIP}({\rm local}) \simeq 70-80\%$, and its total fraction, including dwarfs and elliptical galaxies, is $f_{\rm SNIP}({\rm total}) \simeq 50\%$. { I take two very recent studies of the CSM of SNe Ia, of Tycho's SNR \citep{Kobashietal2023} and of SN 2018evt \citep{WangLetal2023}, to actually support a SNIP scenario for these two SNe Ia. } A point symmetry in a SNR Ia is a natural possibility of the CD scenario when the progenitor PN of a SNIP has a point-symmetry. For a recent study of SNIPs in relation to SNR properties see \cite{Courtetal2023}.  

In the \textit{double degenerate} (DD) scenario (e.g., \citealt{Webbink1984, Iben1984}) without a MED time or with a MED time (e.g., \citealt{LorenAguilar2009, vanKerkwijk2010, Pakmor2013, Levanonetal2015, LevanonSoker2019, Neopaneetal2022}), there is a delay from the end of the CEE to the merger itself due to gravitational wave emission by the double WD system $t_{\rm GW}$. There are several channels of this scenario (e.g., \citealt{Pakmoretal2011, Liuetal2016, Ablimitetal2016,  YungelsonKuranov2017, Zenatietal2019, Peretsetal2019}), with some recent interest in the violent merger channel (e.g., \citealt{AxenNugent2023, Kwoketal2023, Maedaetal2023, Siebertetal2023a, Siebertetaletal2023b, Srivastavetal2023}). 
In the DD scenario, the delay time from the end of the CEE to explosion is $t_{\rm CEED} = t_{\rm GW}$. In the DD-MED scenario, the time from the end of the CEE to the explosion itself includes also the MED time, and therefore $t_{\rm CEED}=t_{\rm GW} + t_{\rm MED}$ (see discussion in \citealt{Soker2022Delay}). The way to form a point-symmetric nebula is if the explosion takes place before the PN material is dispersed into the ISM, i.e., $t_{\rm CEED} \lesssim 10^6 \yr$. However, due to the generally long gravitational-wave merger time $t_{\rm GW}$, this possibility is very rare.  

In the different channels of the \textit{double-detonation}  (DDet) scenario (e.g., \citealt{Woosley1994, LivneArnett1995, Papishetal2015, Shenetal2018a, Shenetal2018b, Ablimit2021, Zingaleetal2023}) the explosion of a CO WD is triggered by the thermonuclear detonation of a helium layer on the WD. This ignition takes place on a dynamic timescale and cannot lead to a point-symmetric morphology. Only if the explosion takes place within hundreds of thousands of years after the CEE of the progenitor binary system, i.e.,  $t_{\rm CEED} \lesssim 10^6 \yr$, this scenario might lead to point-symmetric remnant as being a SNIP.
My estimate \citep{Soker2019Rev}, based in part on the no detection of the surviving companions in SNRs  (e.g., \citealt{LiuDetal2019, Shieldsetal2022, Shieldsetal2023ApJ}), is that the DDet scenario accounts for peculiar SNe Ia (e.g., \citealt{Liuetal2023, PadillaGonzalezetal2023, Yadavallietal2023}), but only rarely for normal SNe Ia. More rare will be normal SNe Ia through this channel that explode before the PNe are dispersed.   

The \textit{single degenerate} (SD) scenario (e.g., \citealt{Whelan1973, HanPodsiadlowski2004, Orio2006, Wangetal2009, MengPodsiadlowski2018, Cuietal2022}) might in principle lead to a point-symmetric SNR if the CSM formed by the wind from a giant mass-donor has a point-symmetric morphology.
This is basically an SN Ia inside a symbiotic nebula. 
Symbiotic progenitors of SNe Ia are very rare (e.g.,  \citealt{LaversveilerGoncalves2023}). 
There are two main differences between symbiotic progenitors and SNIPs. (1) In the case of an SD scenario, the expectation is for the presence of a red giant branch star or an AGB star in the SNR. (2) The CSM mass is much smaller than in a SNIP. The large deceleration of the ejecta of SNR G1.9+0.3 makes this scenario less likely (section \ref{sec:Summary}). 

The very rare (e.g., \citealt{Toonenetal2018, HallakounMaoz2019, HamersThompson2019, GrishinPerets2022}) \textit{WD-WD collision} (WWC) scenario, where two unbound WDs collide with each other (e.g., \citealt{Raskinetal2009, Rosswogetal2009, Kushniretal2013, AznarSiguanetal2014, Glanzetal2023}) does not predict a point-symmetric SNR { as I study here. The collision of two equal mass WDs can lead to a large-scale bipolar structure in case of a head-on collision (e.g., \citealt{Hawley2012}), or to a large-scale point-symmetric ejecta with a very large departure from a large-scale elliptical shape (e.g., \citealt{Glanzetal2023}). The demand for equal-mass WDs in a scenario that is extremely rare to start with and the large departures from an elliptical shape, make this scenario unlikely to explain the point-symmetric morphology of SNR G1.9+0.3 that I study here. }

The overall conclusion from this discussion is that the most likely explanation for a point-symmetric SNR Ia morphology is an SNIP. The scenario that statistically has the largest fraction of SNIPs is the CD scenario. I return to this point in section \ref{sec:Summary}. 

\section{Point-symmetry in SNR G1.9+0.3}
\label{sec:PointSymmetry}
In their recent study \cite{Enokiyaetal2023} combined 26 individual X-ray observations of SNR G1.9+0.3 from 2007-2015 in the energy range of 0.5 to $7 \keV$. They obtained a detailed X-ray image that reveals detailed structures (previous X-ray studies include, e.g., \citealt{Reynoldsetal2008, Reynoldsetal2009, Borkowskietal2010, Borkowskietal2013, Borkowskietal2014,  Borkowskietal2017, Carltonetal2011, Zoglaueretal2015}). In addition, they present contours of molecular emission which they use to identify molecular clouds. In this study, I refer only to the X-ray morphology. I do not consider abundance or molecular clouds. 
 
\cite{Borkowskietal2017} present an X-ray image very similar to that by \cite{Enokiyaetal2023}. The new one allows a better analysis of the point-symmetry. \cite{Borkowskietal2017} present the proper expansion velocities on the plane of the sky and find two strong properties. The first is that the closer to the center arcs on the north and south expand much slower than the ears.
Following \cite{TsebrenkoSoker2015G1903}, I take the arcs to be part of the equatorial structure and the ears to be along the polar directions of the PN into which SNR G1.9+0.3 exploded. 
The second property that \cite{Borkowskietal2017} find is that many regions expand not exactly along radial directions. I attribute these properties of slowly expanding arcs and non-radial expansion directions to the interaction of the ejecta with a non-homogeneous PN shell (the CSM). For such an influence of the CSM on the ejecta it should be massive, $\gtrsim 1 M_\odot$, almost ruling out the SD scenario (see section \ref{sec:PointSymmetry}) where the CSM is due to an AGB wind. \cite{Borkowskietal2014} find that the relative proper expansion rate (percentage per year) of the outer parts of the polar regions (that include the ears) is lower than the inner regions. This indicates substantial deceleration of the outer parts of the ejecta along and near the polar directions, again, requiring a relatively massive CSM. 
{ Some parts in the nebula have expansion velocities that are about half, and even less, than other parts. To decelerate the velocity to half its initial value requires in a momentum-conserving interaction a CSM mass that is about equal to the mass of the ejecta. In an energy-conserving case, there is a need for a larger CSM mass. Overall, the CSM mass should be about equal to the decelerated ejecta mass and more. Since a large fraction of the $\simeq 1.4 M_\odot$ ejecta is decelerated, I estimate the CSM mas to be $\gtrsim 1 M_\odot$.  }

\begin{figure*}[]
	\centering
\includegraphics[trim=0.0cm 8.5cm 0.0cm 2.0cm ,clip, scale=0.80]{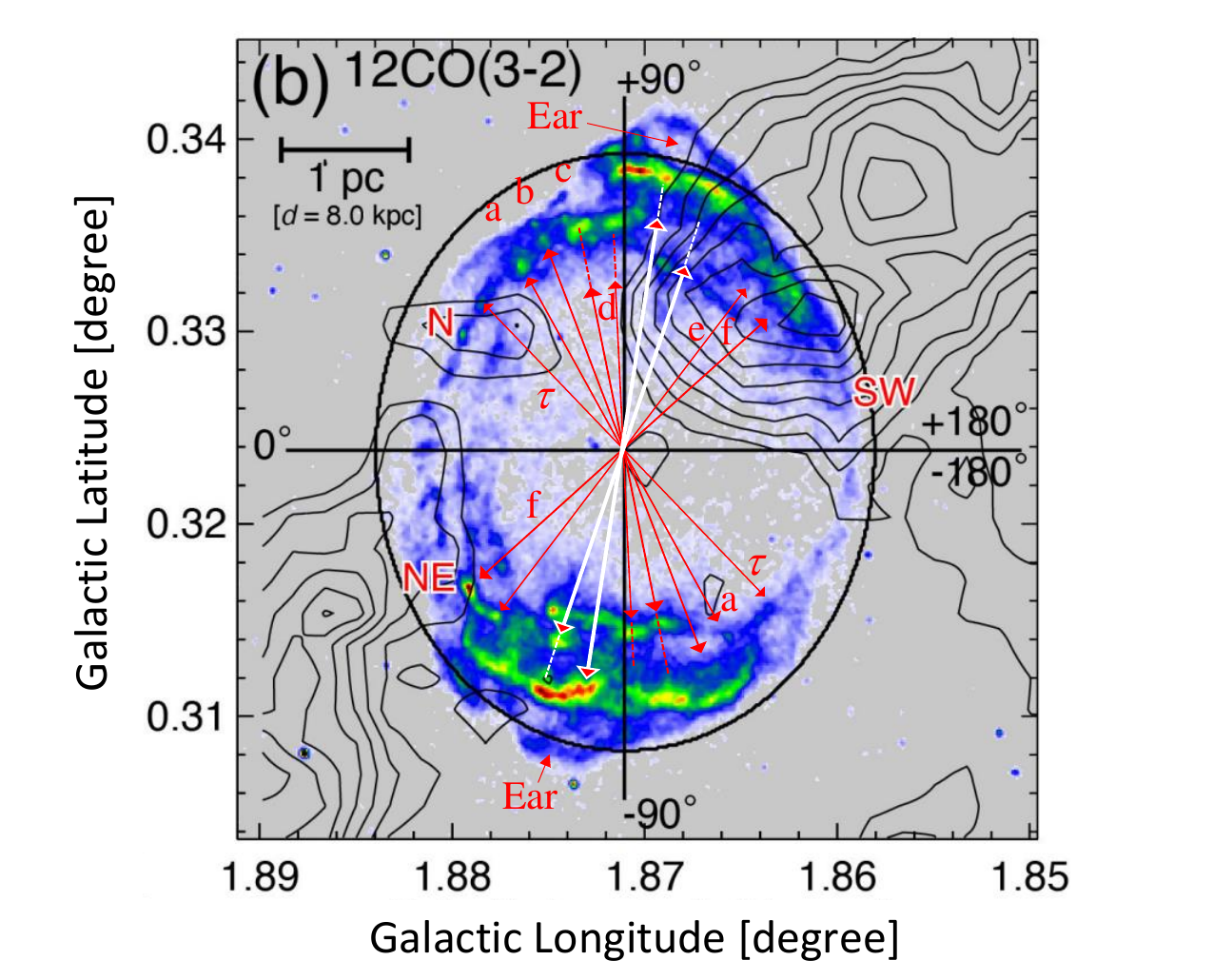}
\caption{An X-ray image with CO contours from \cite{Enokiyaetal2023}. The ellipse and coordinates lines are in the original image. My additions are the double-headed arrows with dashed-line continuations and the marks of the two ears. The center of each double-headed arrow is at the center of the image (where the two black lines cross each other). The six red double-headed arrows DHA-a to DHA-f point at what I interpret as twin clumps of a point symmetric structure, with DHA-$\tau$ indicating a tentative pair due to the small and relatively faint clumps. The two white double-headed arrows signify that although each double-headed arrow points at two opposite clumps to the center, I do not consider them as point-symmetry twins. My interpretation of the point-symmetric structure of these clumps is in Fig. \ref{Fig:SNRfig2}. 
}
\label{Fig:SNRfig1}
\end{figure*}

In Figure \ref{Fig:SNRfig1} I take an image from \cite{Enokiyaetal2023} to which I added the marks of the two ears and added double-headed arrows. I identify six pairs of clumps, marked with double-headed arrows DHA-a to DHA-f, and one tentative, DHA-$\tau$, that form together a point-symmetric structure around the center. I analyze later the two opposite arcs that the two white double-headed arrows point at and reveal a bent point symmetry.    

DHA-e and DHA-f define two opposite arcs at about the same distance from the center. This is the most symmetric point-symmetric component because the two twin arcs (coloured green) are at about the same distance from the center and about the same size. DHA-a points at a clump in the upper part of the image, and at a clump in the bottom that is at about the same distance from the center.  DHA-b points to two clumps along the arrow direction on the upper part of the image, and at a faint green filament at the bottom. Along the direction of DHA-b further away from the faint filament at the bottom, i.e., at the outer part of the SNR, there is the bright arc that DHA-a approximately defines its bright edge. DHA-c and DHA-d point at two twin clumps, but those at the upper part of the image are at a larger distance from the center than the two clumps at the bottom. DHA-d points at a clump in the upper part at the same distance as the bright clump (yellow-red) on the bottom outer arc, as the red-dashed continuation lines show. Additionally to these six pairs, there is a tentative pair marked by DHA-$\tau$. It is tentative because the two opposite clumps are smaller and fainter than the others. 

Overall, the point symmetric structure that the red double-headed arrows define is not perfect although very strong. Considering that the ejecta of this SNR is strongly decelerated, namely interacting with a CSM, it is expected that the point symmetry is not perfect. This is the situation also with tens of PNe (see catalogues listed in section \ref{sec:intro}). 
The asymmetrical interaction of the ejecta of SNR G1.9+0.3 with the CSM and the ISM is evident from the radio images of SNR G1.9+0.3 that present non-uniform brightness and large deviations from spherical symmetry (e.g., \citealt{Greenetal2008, GomezRodriguez2009, Borkowskietal2010, DeHortaetal2014, Borkowskietal2017, Lukenetal2020, Enokiyaetal2023}). As said, this interaction is related also to the non-radial velocity of many parts in this SNR that \cite{Borkowskietal2017} pointed at.   
  
I turn to consider the clumps that the white arrows point at in Figure \ref{Fig:SNRfig1}. Motivated by the bent-morphology of $\approx 10 \%$ of PNe \citep{SokerHadar2002} I consider the same for the two ears of SNR G1.9+0.3 and the bright arc at the base of each ear. In the bent morphology, the symmetry axis is bent through the center, i.e., the angle between the directions to the two opposite clumps/lobes/arc/filaments is $<180^\circ$. In other words, the two opposite structures are displaced in the same direction perpendicular to the symmetry axis. 

\begin{figure*}[]
	\centering
\includegraphics[trim=0.0cm 8.5cm 0.0cm 2.0cm ,clip, scale=0.80]{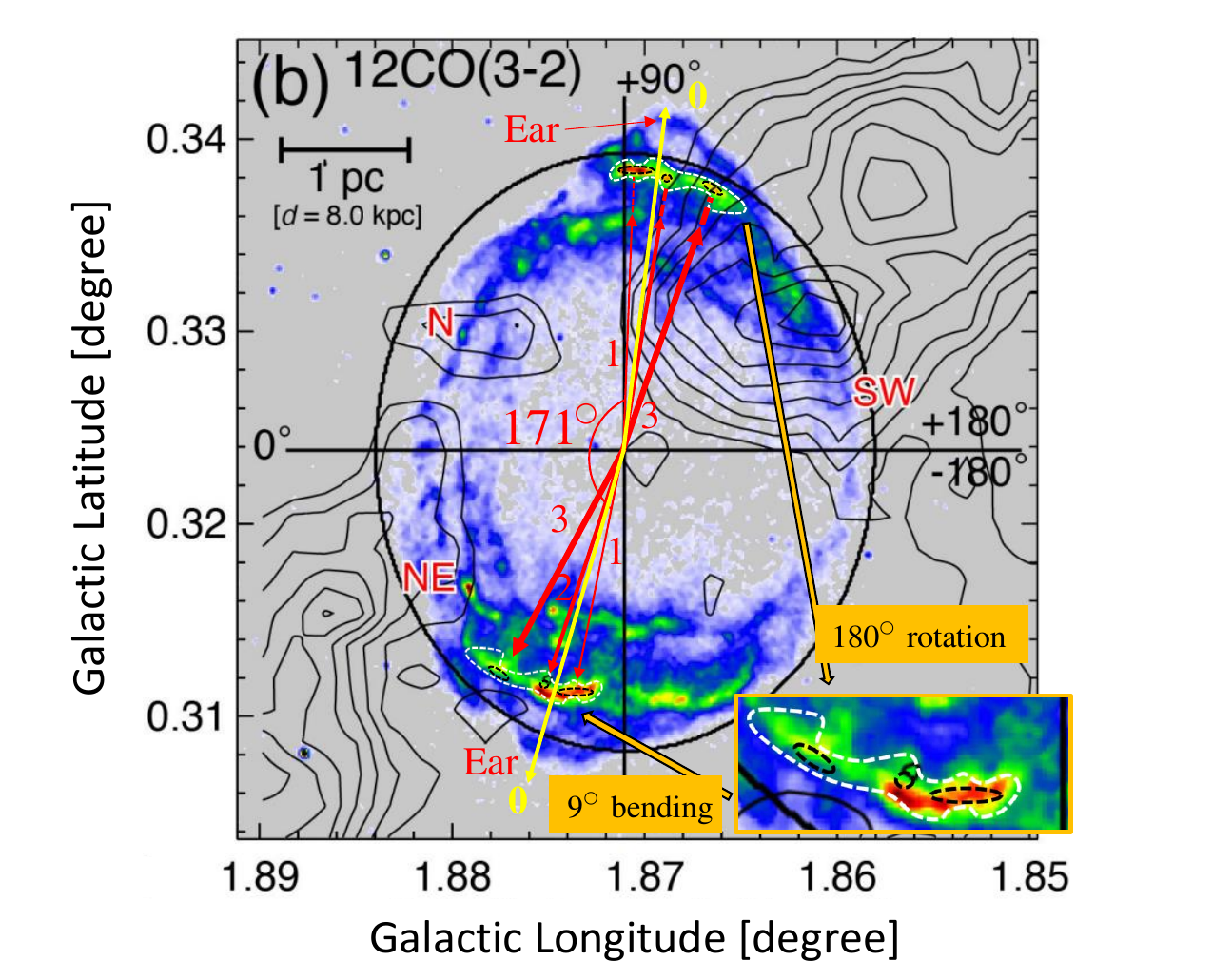} 
\caption{Presentation of the bent point-symmetrical structure of SNR G1.9+0.3. The original X-ray image from \cite{Enokiyaetal2023} is the same as in Figure \ref{Fig:SNRfig1}. I marked the arc at the base of the upper (western) ear with a dashed-white line and its three peaks (yellow-red) with three dashed-black lines. DHA-1 to DHA-3 point at these clumps.  
I copied and rotated this structure around itself by $180^\circ$ and matched it to the arc at the base of the bottom (eastern) ear. The inset on the lower right enlarges this region. There is a $9^\circ$ bent point symmetry of the two ears (DHA-0) and of the two base arcs (DHA-1 to DHA-3).  
} 
\label{Fig:SNRfig2}
\end{figure*}

In Figure \ref{Fig:SNRfig2} I present the $9^\circ$-bent morphological feature of the ears of SNR G1.9+0.3. I construct it as follows. 
I circle the green-coloured arc at the base of the upper (western) ear with a dashed-white line. I also circle by dashed-black lines the three red-yellow peaks inside this arc. I then copy this entire structure to the bottom (eastern) ear and rotate it around itself by $180^\circ$ and displace it to match the arc at the base of the eastern ear. I enlarge the bottom (eastern) arc in the inset on the lower-right of Figure \ref{Fig:SNRfig2}. I find that the best match of the two twin arcs is when the angle through the center is $171^\circ$ instead of $180^\circ$ as marked on Figure \ref{Fig:SNRfig2}. I also added to the figure two yellow arrows at $171^\circ$ to each other, each arrow through the tip of an ear. The four bent double-headed arrows in Figure \ref{Fig:SNRfig2} define the $9^\circ$-bent point-symmetrical morphological component of SNR G1.9+0.3. 

{ Based on the classification of bent-morphology planetary nebulae, I consider the value of $9^\circ$ bent to be significant. For example, the planetary nebula NGC 6826 is classified to have a bent morphology \citep{SokerHadar2002} although its bending angle is only $7^\circ$. The features on which I based the bent morphology are bright, namely, two opposite tangential arcs (marked by dashed-white lines), with bright clumps inside each of the two arcs. Overall, I consider the bent morphology to be observationally significant. }

I note that \cite{Chiotellisetal2021} consider the ears to form in the equatorial plane. This cannot account for a point-symmetry near the ears as I find here. The point-symmetry that I identify in SNR G1.9+0.3 shows in very strong terms that the ears are along the polar directions  (e.g., \citealt{TsebrenkoSoker2013, TsebrenkoSoker2015G1903}) and not in the equatorial plane. Most likely jets shaped the point-symmetrical structure of SNR G1.9+0.3 through their shaping of a PN shell. This brings me to discuss this SNR as an SNIP. 

\section{Discussion and summary}
\label{sec:Summary}

In this short study I analyzed a new X-ray image of  SNR G1.9+0.3 \citep{Enokiyaetal2023} and revealed a clear point-symmetric morphology. I now discuss the possible implications on the SN Ia scenario that best explains the youngest SN Ia in the Galaxy. 

Figures \ref{Fig:SNRfig1} and \ref{Fig:SNRfig2} present the point-symmetric structural features (the point-symmetric morphology) that I identify in SNR G1.9+0.3. In addition to the ears, there are several pairs of clumps and arcs that I identify. In several pairs and one tentative pair the two twin clumps/arcs are on opposite directions, sometimes at somewhat different distances from the center (Figure \ref{Fig:SNRfig1}). The ears and the arc at the base of each ear form a bent point-symmetrical structure, as mark by DHA-0 to DHA-3 on Figure \ref{Fig:SNRfig2}. 

The point-symmetric structure that I identify in SNR G1.9+0.3 is composed of opposite pairs of clumps/arcs/ears that have different directions (the directions of the double-headed arrows). Opposite pairs of jets with varying axis directions, like due to precession, form such structures in a rich variety of astrophysical systems, e.g., from PNe to jet-shaped bubbles in clusters of galaxies. Since explosion models of SNe Ia do not have jets with varying directions (section \ref{sec:Scenarios}), the most likely explanation is that the ejecta of SNR G1.9+0.3 expands into a point-symmetric CSM. The substantial deceleration of the ejecta of SNR G1.9+0.3 requires a massive CSM, which is more likely to be a PN that was expelled during a CEE in the CD scenario than an AGB wind in the SD scenario (section \ref{sec:Scenarios}). Although the DD and the DDet scenarios might also occur shortly after the CEE, the probability for that is much lower than that in the CD scenario (section \ref{sec:Scenarios}).  
{ Also, based on the upper bound of its $^{44}$Ti abundance \cite{Kosakowski2023} argue that SNR G1.9+0.3 is most consistent with a near-$M_{\rm Ch}$ progenitor. The CD scenario is compatible with that finding. }

The interaction of the ejecta with the PN started some tens of years ago at a radius of $\la 1 \pc$. PNe can have such sizes, e.g., the PN IPHASX J055226.2+323724 in the open cluster  M37 \citep{Fragkouetal2022} with an age of $\simeq 10^5 \yr$, (\citealt{Fragkouetal2022, Werneretal2023}). Therefore, the explosion could have taken place while the PN was still shining, rather than explosion into an old post-PN shell. 

I conclude that the most likely explanation for the point-symmetry SNR G1.9+0.3 is an SNIP where the explosion took place into a PN (rather than a remnant of a PN). The explosion destroyed the WD, hence destroyed the PN.

\section*{Acknowledgments}
{ I thank an anonymous referee for helpful comments. } 
This research was supported by a grant from the Israel Science Foundation (769/20).




\label{lastpage}

\end{document}